\documentclass{elsart}
\usepackage{epsfig}
\usepackage{graphicx}
\usepackage{amssymb}

\newcommand{\adag}{a^{\dag}}
\newcommand{\bdag}{b^{\dag}}

\begin{document}
\begin{frontmatter}

\title{Intermediate statistics as a consequence of deformed algebra}

\author{A Lavagno$^{1,2}$ and P Narayana Swamy$^3$}
\address{$^1$Dipartimento di Fisica, Politecnico di Torino, I-10129, Italy.}
\address{$^2$Istituto Nazionale di Fisica Nucleare (INFN),
Sezione di Torino, I-10126, Italy.}
\address{$^3$Department of Physics, Southern Illinois
University, Edwardsville, IL 62026, U.S.A.}

\maketitle

\begin {abstract}
We present a formulation of the deformed oscillator algebra which
leads to intermediate statistics as a continuous interpolation
between the Bose-Einstein and Fermi-Dirac statistics. It is
deduced that a generalized permutation or exchange symmetry leads
to the introduction of the basic number and it is then established
that this in turn leads to the deformed algebra of oscillators. We
obtain the mean occupation number describing the particles obeying
intermediate statistics which thus establishes the interpolating
statistics and describe boson like and fermion like particles
obeying intermediate statistics. We also obtain an expression for
the mean occupation number in terms of an infinite continued
fraction, thus clarifying successive approximations.
\end {abstract}


\end{frontmatter}

\section {Introduction}

The intermediate statistics that continuously interpolates between
the Bose-Einstein statistics (BE) and the Fermi-Dirac statistics
(FD) is applicable to some particles often referred to as anyons
\cite{Leinnas}. The ensemble of such particles has been of great
interest in the study of many topics such as Chern-Simons
\cite{Chern} gauge theory of fields, Aharonov-Bohm effect
\cite{Aharanov-Bohm}, the second virial coefficient, the braid
group \cite{Lerda}, fractional statistics and anyon
superconductivity \cite{Wilczek}. The study by Arovas et al. in
Ref.\cite{Arovas} arising from an exact two-particle partition
function leads to a computation of the second virial coefficient
valid for low densities and/or  high temperatures. The work in
Ref.\cite{RA-PNS1} outlines the consequences of an appropriately
approximate distribution function for the anyons (intermediate
statistics) and examines various properties of  the
thermostatistics of interpolating statistics such as the partition
function, entropy, pressure, internal energy, the various virial
coefficients and the specific heat of anyons compared to the
properties of standard bosons or fermions.


More recently, the same subject has been examined from the point of
view of detailed balance \cite{RA-PNS2}. Employing standard methods
of statistical physics, the theory of exchange symmetry has been
shown in this work to lead to a continuous interpolation between
Bose and Fermi statistics. The basic numbers arise naturally in this
theory as a consequence of exchange or permutation symmetry. This
theory has many interesting features, one of them being that this
formulation is not restricted to 2 + 1 space-time dimensions. Indeed
such a formulation leads to a mean occupation number expressed as an
infinite continued fraction so that the meaning of successive
approximations is well clarified.


There is another generalization of the standard thermostatistics
that has been studied extensively in the literature and this has to
do with the theory of $q$-deformed quantum oscillators or quantum
groups \cite{Biedenharn,Macfarlane}.
It has been shown that a complete formulation of generalized
thermostatistics of $q$-bosons and $q$-fermions can be established
using basic numbers with the base $q$ and employing the
$q$-calculus based on the Jackson Derivative (JD). The
thermodynamic functions such as entropy, pressure, internal
energy, specific heat etc. of such deformed systems have been
studied and compared with standard bosons and fermions
\cite{PNS,ALPNS1,ALPNS2,ass,jpa2008,pla2002,gkqua,albe,algin,su}.
The method of detailed balance may be employed for the purpose of
establishing an intermediate statistics and this is known to
require the use of basic numbers or bracket numbers. Such bracket
numbers arise naturally in the $q$-deformed algebra of harmonic
oscillators in such a manner that the limit of $q\rightarrow 1$
corresponds to the boson and fermion oscillators. However, the
task of establishing the equivalence of a deformed algebra and the
description of intermediate statistics has been an outstanding
problem. A basic problem, as has been pointed out in the
literature \cite{Frappat,Chaichian} is that the oscillator algebra
is valid in any number of dimensions while anyons arise only in
two space dimensions. However, it is quite possible that
intermediate statistics may indeed be different from that of
anyons. The legitimate question accordingly arises naturally: what
kind of deformation can successfully describe interpolating
statistics?

 In the present work, we shall seek  a fresh attempt at
formulating the intermediate statistics in the framework of a
$q$-deformed oscillator algebra. We shall present the basic
formulation to illustrate the  equivalence between intermediate
statistics and the deformed algebra of oscillators.  Due to the
special issues arising in two dimensions, we shall restrict our
formulation to the intermediate statistics applicable to particles
which may indeed differ from the anyons in the literature. Section
2 introduces the idea of a basic number, i.e., we establish the
fact that a general kind of permutation implies the existence of
the basic number. We believe that this demonstration has so far
been an outstanding problem in the literature. We devote Section 3
to the demonstration that the basic number naturally implies a
deformed algebra. We study the intermediate statistics of
particles in Section 4, specifically the mean occupation number
for ensembles of such particles. We also briefly study boson like
and fermion like intermediate statistics here. As for the
consequences in thermostatistics of such a deformation, we shall
simply refer to earlier work \cite{RA-PNS1} which is based on the
first approximation of the occupation number, while it must be
pointed out that the occupation number dealt with in the present
work is exact. We illustrate the nature of approximations by
expressing the mean occupation number in intermediate statistics
as an infinite continued fraction in Section 5. We present a brief
summary in Section 6.

 \section{Deformed $\kappa$-permutation implies existence of the basic
number}


We need to briefly summarize the procedure for dealing with many
particles and the quantum probabilities, according to the method
developed by Feynman \cite{Feynman}. Feynman's method pays
attention to the special rules for the interference that occurs in
processes with identical particles and considers in detail the
direct and exchange amplitudes. Such attention is necessary for
the determination of the enhancement factor of $(N+1)$ particles
for bosons, which measures the probability of an additional boson
from a state of n boson particles. We proceed as follows.


 The operation of permutation or exchange of the
coordinates of $N$ particle wave function results in
multiplication by the statistics determining parameter $\kappa$,
which may be  a complex number in general,  with the property
$|\kappa|^2=1$. In the following, for simplicity, we will limit
ourselves to consider spinless particle although the results can
be easily extended to the general case.  We shall begin with the
two particle state wave function defined in terms of single
particle wave functions
\begin{equation}\label{2}
    \psi^{(2)}= \frac{1}{\sqrt{N_2}}(\psi_a(r_1) \psi_b(r_2) + \kappa \;
 \psi_a(r_2) \psi_b(r_1) )\, ,
\end{equation}
and the corresponding probability density associated with the two
particle state:
\begin{eqnarray}\label{2a}
P^{(2)}= \frac{1}{N_2}\; &&(\psi_a(r_1) \psi_b(r_2) + \kappa \;
 \psi_a(r_2) \psi_b(r_1) )^* \times \nonumber \\
 &&(\psi_a(r_1) \psi_b(r_2)+ \kappa \;
 \psi_a(r_2) \psi_b(r_1) )\, ,
\end{eqnarray}
according to Feynman's method. Here, $N_2=1+|\kappa|^2=2$ for the
two particle state.

In the case of identical particles, the process in with the
particle with quantum numbers $a$ going into $r_1$ and $b$ going
into $r_2$ cannot be distinguished from the exchange process in
which $a$ goes into $r_2$ and $b$ goes into $r_1$. Thus, for
simplicity, we set in the following $r_1=r_2$ and the probability
density for identical particles reduces to
\begin{equation}\label{3}
    P^{(2)}\Longrightarrow \frac{1}{N_2}
(1 + \kappa^{-1} )(1+\kappa) \, \Pi_2\, ,
\end{equation}
where $\Pi_2=|\psi_a|^2 |\psi_b|^2$.

 We can thus express the probability density associated with the two particle
state in the form
\begin{eqnarray}\label{4}
P^{(2)}&=& \half (1+\kappa^{-1})(1 + \kappa)\,\Pi_2=
\half(1+1+\kappa+\kappa^{-1})\,\Pi_2\nonumber\\
&=&\half (2 + [2])\,\Pi_2\, ,
\end{eqnarray}
where we have introduced the basic number or bracket number $[x]$
in the last step \cite{Exton}. We shall employ the basic numbers
in the symmetric formulation, accordingly defined by the general
form:
\begin{equation}\label{5}
[x]=\frac{\kappa^x-\kappa^{-x}}{\kappa-\kappa^{-1}}\, ,
\end{equation}
Concomitantly, we can introduce the operator form
\begin{equation}\label{6}
 [N]=\frac{\kappa^N- \kappa^{-N}}{\kappa - \kappa^{-1}}\, ,
\end{equation}
where $[N]$ is not the number operator but $N$ is. It is
well-known that the basic number $[n]$ defined above is the sum of
the geometric series and thus we may also frequently employ the
series form
\begin{equation}\label{7}
    [n]=\kappa^{-n+1}+ \kappa^{-n+3} + \cdots + \kappa^{n-3}+
    \kappa^{n-1}\, .
\end{equation}
The parameter $\kappa$  is in general a complex number, and at
this point we may express it as  $\kappa = e^{i \alpha}$ where
$\alpha$ is a real number. The statistics determining parameter
takes the values $\kappa = \pm 1$ corresponding to $\alpha = 0,
\pi$ which define the Bose-Einstein (BE) and Fermi-Dirac (FD)
cases respectively. The parameter has the properties $\kappa^* =
\kappa^{-1}, \; |\kappa|=1$.


 The above form for the probability, as in Eq.(\ref{4}) associated with
the two particle state is correct since $[2]= \kappa + \kappa^{-1}$
is true. It is evident that the probability is real as stated in
Eq.(\ref{4}), as will be observed to be the case for probabilities
of multiparticle states. The wave function, Eq.(\ref{2}) contains
both the direct and the exchange terms, and it is the exchange term
which contains the factor $\kappa$. In the limit when $\kappa
\rightarrow 1$ it reduces to the standard Bose case and reduces to
the Fermi case when $\kappa \rightarrow -1.$


 Next we consider the three particle state defined by
\begin{eqnarray}\label{8}
\psi^{(3)}= \frac{1}{\sqrt{N_3}}(&&a_1 b_2 c_3 + \kappa \; a_1 b_3
c_2+
    \kappa\; a_2 b_1  c_3  \nonumber \\
    &&+\kappa^2 \; a_3 b_1 c_2  + \kappa^2 \; a_2 b_3 c_1+
    \kappa^3 \; a_3 b_2 c_1  )\,
   ,
\end{eqnarray}
where we have used the abbreviation $a_1=\psi_a(r_1), \;
b_2=\psi_b(r_2), \; c_3=\psi_c(r_3)$ in the first term of the
bracket and consequent permutation of the indexes in the following
terms. In this case we observe that the normalization is given by
$N_3=1 + 2 |\kappa|^2+ 2 |\kappa|^4+ |\kappa|^6= 1 + 2 + 2 + 1 = 6
= 3!$. We may now compute the probability density for the state to
be
\begin{equation}\label{9}
P^{(3)}= \frac{1}{6}(a_1 b_2 c_3 + \kappa\;  a_1 b_3 c_2+
...)^*(a_1 b_2 c_3 + \kappa \; a_1 b_3 c_2+ ...)\, .
\end{equation}

Specializing to the case of identical particles (and setting as
before $r_1=r_2=r_3$), the probability density reduces to
\begin{equation}\label{10}
P^{(3)}= \frac{1}{6}(1+2\kappa^{-1}+2 \kappa^{-2}+\kappa^{-3})
(1+2\kappa +2 \kappa^{2}+\kappa^{3})\, \Pi_3\, ,
\end{equation}
where, analogously as before, we have defined $\Pi_3=|a|^2 |b|^2
|c|^2$. Again, we write in terms of the basic numbers by using,
$[2]=\kappa + \kappa^{-1}, \; [3]=\kappa^2+\kappa^{-2}+1, \;
[4]=\kappa^3 +\kappa +\kappa^{-1}+\kappa^{-3}$ etc. with $[n]$
given by Eq.(\ref{7}). We may thus re-express the above
probability density as
\begin{equation}\label{11}
    P^{(3)}= \frac{1}{6} (6 + 7 [2] + 4[3]+ [4])\, \Pi_3\, ,
\end{equation}
  and we may also express it in the convenient and useful form
\begin{equation}\label{12}
    P^{(3)}= \frac{1}{6}(2 + [2])(2 + 2[2]+[3])\, \Pi_3\, .
\end{equation}
To establish the equivalence of the above two forms, we first need
to obtain some formulae.
 Consider the sum
\begin{equation}\label{13}
    [1] + [3] + [5] + \cdots + [2n-1]\, .
\end{equation}
Since $\kappa = e^{i \alpha}$, we may express the basic number in
terms of the trigonometric functions
\begin{equation}\label{14}
    [n]= \frac{\kappa^n - \kappa^{-n}}{\kappa - \kappa^{-1}}
    =\frac{\sin n\alpha}{ \sin \alpha}\, .
\end{equation}
Let us remark that, although the parameter $\kappa$ can be in
general a complex number, the basic number $[n]$ is a real number
(if $n$ is a real number).

The above series may accordingly be written as
\begin{equation}\label{15}
\hspace{-10mm}[1] + [3] + [5] + \cdots + [2n-1]=\frac{\sin \alpha
+ \sin 3\alpha + \cdots + \sin (2n-1)\alpha}{\sin \alpha}\, .
\end{equation}
We can now sum the series of the sine functions by employing
 the  identity \cite{GR}:
\begin{equation}\label{16}
    \sum_{k=1}^n\; \sin(2k-1)\alpha = \frac{\sin^2 n\alpha}{\sin
    \alpha}\, .
\end{equation}
Thus we obtain
\begin{equation}\label{17}
[1] + [3] + [5] + \cdots + [2n-1]= \frac{\sin^2 n\alpha}{\sin^2
\alpha}= [n][n]\, ,
\end{equation}
and hence  we derive the important formula:
\begin{equation}\label{18}
[n]\; [n]= [1] + [3] + [5] + \cdots + [2n-1]\, .
\end{equation}
Employing this formula, we see that the form of the probability
density $P^{(3)}$ in Eq.(\ref{11}) and Eq.(\ref{12}) are clearly
the same. In a similar manner, we can also establish the following
result:
 \begin{equation}\label{19}
  [1][2] =[2],
  [2][3] = [2] + [4],
  [3][4] = [2] + [4] + [6], \cdots \, ,
  \end{equation}
and more generally for any $n$:
\begin{equation}\label{20}
  [n-1][n] = [2] + [4] + [6] + \cdots + [2(n-1)]\, .
\end{equation}

 Next we can perform the same exercise for the
four-particle state. After much algebra, we determine the
probability density associated with the four particle states ($a$,
$b$, $c$, $d$) to be
\begin{equation}\label{24}
    P^{(4)}=\frac{1}{4!}\; (2+[2])(2+2[2]+[3])(2+2[2]+2[3]+[4])\, \Pi_4\, ,
\end{equation}
where $\Pi_4=|a|^2 |b|^2 |c|^2 |d|^2$.

This enables us to generalize to the case of $N$-particle states
($a$, $b$, $c$, $\cdots$). Thus we finally determine
\begin{eqnarray}
  P^{(N)}=\frac{1}{N!}\;
    &(& 2+[2])(2+2[2]+[3])(2+2[2]+2[3]+[4])\cdots  \nonumber\\
   &(&2 + 2[2]+ 2[3]+ \cdots + 2[N-1]+[N])\, \Pi_N\, ,
   \label{25}
\end{eqnarray}
where $\Pi_N=|a|^2 |b|^2 |c|^2 \cdots$  and we have thus obtained
the probability density for the $N$-particle state in the desired
closed form.


Finally, let us consider the BE and FD limits. If we take the
limit $\kappa \rightarrow +1$, we find $P^{(2)}\rightarrow
2!\,\Pi_2$; $P^{(3)}\rightarrow 3!\,\Pi_3$; $P^{(4)} \rightarrow
4!\, \Pi_4$; $\cdots$; $P^{(N)} \rightarrow N!\,\Pi_N$, thus
obtaining the true BE limit. Furthermore we also confirm the fact
that in the BE limit, the enhancement factor determined by
$P^{(N+1)}/P^{(N)}$ turns out to be proportional to $N+1$, as it
should exactly be. Now let us consider the Fermi limit when
$\kappa \rightarrow -1$. We know that $[2] \rightarrow -2$ in this
limit. Hence $2 + [2]$ vanishes. Since this is the common factor
for the general probability density $P^{(n)}$, we conclude that
all the probabilities vanishes for $n=2,3, \cdots N$ in the Fermi
limit, which is in accordance with the Pauli exclusion principle
for standard fermions. In this manner, we are in a sense
confirming the correctness of all the analysis above.


Although we may not explicitly need the closed form for the
probabilities, it is important to recognize that the
$\kappa$-deformed permutation or exchange of particles is strictly
related to the existence of the basic number.
In the following Section, we will establish the connection between
the basic number, in the symmetric formulation, and the deformed
algebra.


Finally, let us mention that there arises an interesting situation
in the case of two spatial dimensions, due to the fact that
permutation is equivalent to a rotation followed by a translation.
The case of such an exchange in two dimensional space corresponds
to the symmetry of the braid group \cite{Lerda} but we shall not
concern ourselves with this feature at this time.


\section{ Basic number implies the deformed algebra}


 Let us begin from the definition of the basic number
in the case of  particles obeying intermediate statistics, given
in Eq.(\ref{6}), and since the Fock state $|n\rangle$ is an
eigenstate of the number operator $N$, we also have
\begin{equation}\label{26a}
[N]|n\rangle = [n]|n \rangle,\;\;
     [n]=\frac{\kappa^n-\kappa^{-n}}{\kappa
    -\kappa^{-1}}\, ,
\end{equation}
 where $n=0,1,2,\cdots .$ It is readily seen from the definition that the  basic
numbers satisfy the relation
\begin{equation}\label{27}
    [n+1]=\kappa [n]+ \kappa^{-n}\, .
\end{equation}
We may now identify the operator $[N]$ in terms of the creation
and annihilation operators by the definition:
\begin{equation}\label{28}
    [N]=\adag a\, ,
\end{equation}
where it is expected that the creation operator raises the number
of quanta and the annihilation operator lowers the number of the
quanta in the Fock state representation, so that we may express
the raising and lowering property as
\begin{equation}\label{29}
    \adag |n\rangle = C_n'|n+1 \rangle, \; a |n\rangle = C_n|n-1
    \rangle\, ,
\end{equation}
where $C_n, \; C_n'$ are constants to be determined. We observe that
Eq.(\ref{28}) is our second assumption. This last assumption may not
truly or necessarily be required since $[N]$ must go to $N$ in the
limit when $\kappa \rightarrow 1$. The constants $C_n, \; C_n'$ are
determined in the customary manner as
\begin{equation}\label{30}
    C_n= \sqrt{[n]}, \;\; C_n'= \sqrt{[n+1]}\, .
\end{equation}
Next we obtain the relations
\begin{equation}\label{32}
    [N]\adag |n\rangle =[N]
    C_n'|n+1\rangle = [n+1]C_n'|n+1\rangle, \;
\end{equation}
and
\begin{equation}\label{33}
    \adag [N] |n\rangle =\adag [n]|n\rangle =[n]\adag |n\rangle=
 [n] C_n'|n+1\rangle\,,
\end{equation}
in the derivation of which,  we notice with care that eigenvalues
commute, while operators in general need not. We also note that
the above relation is true for any state $|n\rangle$.


 From this we derive the relations, valid for any $|n\rangle$,
 \begin{equation}\label{34}
    ([N]\adag - \kappa \adag [N])|n \rangle = \adag \kappa^{-N}|n
    \rangle \, ,
\end{equation}
which can be expressed as an operator identity:
\begin{equation}\label{35}
([N]\adag - \kappa \adag [N]) = \adag \kappa^{-N}\, .
\end{equation}
If we employ the definition, Eq.(\ref{28}),  the above can be
written in the form
\begin{equation}\label{36}
    \adag a \adag - \kappa \adag \adag a=\adag \kappa^{-N}\, ,
\end{equation}
noting that this is an operator relation.  From this, the deformed
algebra follows immediately:
\begin{equation}\label{37}
    a \adag   - \kappa \adag a =\kappa^{-N}\, ,
\end{equation}
which is an operator identity.


 Hence we have established the connection between the
basic number and the deformed algebra i.e., the basic number
introduced in Section 2 implies that the creation and annihilation
operators obey the deformed algebra.


 At this point,  we may make a general remark in light of  our
formulation. What we have presented is the algebra describing
deformed oscillators. Thus we infer that the three cases $+1, -1$,
and $\kappa $ corresponding to BE, FD, and intermediate statistics
respectively, are in one-to-one correspondence  with the
oscillator algebras
    as follows:
\begin{equation}\label{38}
a \adag - \adag a = 1; \;\; b \bdag + \bdag b =1; \;\; a \adag -
\kappa \; \adag a =\kappa^{-N} \, ,
    \end{equation}
where the variable parameter $\kappa$ may be complex in general
with end points (limits) -1 and +1. Let us observe that the above
third equation is formally equivalent to symmetric $q$-deformed
oscillator algebra \cite{ALPNS2,Chaichian}. This might be asserted
as our basic premise.

 We are merely attempting to point out  the above
correspondence. We just note that this is analogous to the
Bargmann-Wigner holomorphic representation \cite{Bargmann}.
However, the pertinence of the connection with the two dimensional
case is immediately evident. Planar physical systems, in two space
and one time dimensions, display many peculiar and interesting
quantum properties owing to the unusual structure of rotation,
Lorentz and Poincare groups and thus lead to a theory of anyons.
That is, in the case of two dimensions, as pointed out extensively
by Ref.\cite{Lerda}, the exchange or permutation of two objects is
described by the braid group with its direct consequences. In the
present work we do not explicitly deal with the case of two space
dimensions. We consider only the deformed algebra without
resorting to any specific dimensions and demonstrate that the
deformation of the algebra, derived form the properties of the
basic numbers and strictly connected to $\kappa$-permutations, can
lead to intermediate statistics with a deeper physical insight
with respect to the standard formulation of the $q$-deformed
oscillators.

Now we proceed to study the consequence of the deformed algebra
for the particles of intermediate statistics.

\section{ Intermediate statistics: mean occupation number}

Let us start from the following Hamiltonian
\begin{equation}\label{42}
H=\sum_i N_i (E_i-\mu)\, ,
\end{equation}
It is important to recognize that the latter Hamiltonian, in spite
of the appearance, does include deformation, as will become
evident from the form of the average occupation number.

Next we study the mean value of the occupation number in the
standard manner.
 We begin by introducing the mean value, after omitting the
 subscript for convenience,
\begin{equation}\label{43}
\kappa^{n}=\frac{1}{Z} \, Tr (e^{-\beta H } \kappa^{N})\,.
\end{equation}
 Next, we consider the expectation value
\begin{equation}\label{44}
[n] =  \frac{1}{Z}\, Tr (e^{-\beta \sum(E_i-\mu)(N_i+1) }a \adag )
\, ,
\end{equation}
where $n$ is a real number that represents the mean occupational
number of particles.

We may now use the cyclic property of the trace,  and the relation
$a f(N) = f(N+1) a$, which can be established in a straightforward
manner for any polynomial function $f(N)$, in order to deal with
the right hand side above.  Proceeding in the standard manner, we
obtain
\begin{equation}\label{45}
 \frac{[n]}{[n+1]}=e^{-\eta}\,,
\end{equation}
where $\eta= \beta (E-\mu)$. This is an important step. Proceeding
further, we obtain, after some algebra,
\begin{equation}
 \label{46}
 [n]=\frac{\kappa^{-n}}{e^{\eta }- \kappa}\,.
\end{equation}
We may now proceed to determine the occupation number $n$. From
the above equation, we obtain
\begin{equation}\label{47}
    (\kappa^2)^n= \frac{e^{\eta}-\kappa^{-1}}{e^{\eta}-\kappa}\, ,
\end{equation}
which leads to the result
\begin{equation}\label{48}
    n= \frac{1}{2 \ln \kappa} \ln \left( \frac{e^{\eta}-
    \kappa^{-1}}{e^{\eta}-\kappa} \right)\, ,
\end{equation}
which expresses the form of the mean occupation number for the
particles obeying interpolating statistics. It is now evident that
the chosen Hamiltonian does indeed include the deformation.
Furthermore, let us point out that, although the parameter
$\kappa$ can be in general a complex number, the mean occupational
number $n$ expressed in Eq.(\ref{48}) is effectively a real
number. In fact, as before, if we write $\kappa=\exp(i\,\alpha)$
where $\alpha$ is a real number, it is easy to show that the
factor in parenthesis of Eq.(\ref{48}) can be written as
$(e^{\eta}-\kappa^{-1})/(e^{\eta}-\kappa)=\exp(i\,2n\alpha)$.

In this context, it is relevant to observe that the statistical
origin of such $q$-deformation lies in the modification, relative
to the standard case, of the number of states $W$ of the system
corresponding to the set of occupational number ${n_i}$
\cite{ALPNS1}. In the subject literature, other statistical
generalizations are present, such as the so-called nonextensive
thermostatistics or superstatistics with a completely different
origin \cite{tsallis,book1,book2,varie,abe1,beck}.

 Now we are going to further express the occupation
number in an approximate form. At this scope, for simplicity, in
the following we limit our considerations by requiring the
parameter of the deformation to be a real number, more explicitly,
taking the projection of $\kappa$ into the real axis. Therefore,
hereafter we will set $\overline{\kappa}={\rm Re} [\kappa]$.

From Eq.(\ref{48}), we arrive at the result in the form of a power
series
\begin{eqnarray}
  n=&&\frac{1}{y}
   +\left(\frac{1}{3\,y^3}+\frac{\overline{\kappa}}{2\,y^2}+\frac{1}{6\,y}\right)\,\epsilon^2+
    \left(\frac{1}{3\,y^3}+\frac{\overline{\kappa}}{2\,y^2}+\frac{1}{6\,y}\right)\,\epsilon^3\nonumber\\
  &&\left(\frac{1}{5\,y^5}+\frac{\overline{\kappa}}{2\,y^4}+\frac{13}{18\,y^3}+
  \frac{7\,\overline{\kappa}}{12\,y^2}+\frac{29}{180\,y}\right)
    \,\epsilon^4   + \cdots\label{49} \, ,
\end{eqnarray}
where $y= e^{\eta}- \overline{\kappa}$ and we have taken
$\overline{\kappa} = 1- \epsilon, \; \epsilon \ll 1$. At this
point, if we use the approximation by retaining only the leading
term  we then arrive at the form
\begin{equation}\label{50}
    n\approx \frac{1}{e^{\eta}- \overline{\kappa}}\, .
\end{equation}
This is the  form familiar from the work of Ref.\cite{RA-PNS1}
which, albeit the approximation, contains many interesting
applications. More generally, for $\epsilon$ not very small, we have
the series form in Eq.(\ref{49}) describing the various powers of
the deformation parameter $\epsilon$.


It might now appear as though we may have obtained the above
result for both the boson B-type and the fermion F-type
intermediate statistics. Accordingly, we need to examine the
B-type and F-type cases separately.


 Let us first consider the B-type case of
intermediate statistics. The algebra is as before, with the
occupation number given in Eq.(\ref{48}), but the range may now be
taken as $0 \leq \overline{\kappa} \leq 1$ with the understanding
that $\overline{\kappa} =1$ corresponds to the BE case. The series
form for $n$ is given by Eq.(\ref{49}) and the approximate form is
given by Eq.(\ref{50}). Ref. \cite{RA-PNS1} provides many
applications of the B-type intermediate statistics.


We may now consider the F-type particles.  Now the range for the
F-type particles is $-1 \leq \overline{\kappa} \leq 0$. Setting
$\overline{\kappa} = - \lambda$, we may modify Eq.(\ref{47}) as
\begin{equation}\label{52}
    (\lambda^2)^n= \frac{e^{\eta}+\lambda^{-1}}{e^{\eta}+\lambda}\, .
\end{equation}
We may express the range in the form $0 \leq \lambda \leq 1$. The
algebra above  leads to the relation
\begin{equation}\label{53}
    n=\frac{1}{2 \ln \lambda}\; \ln(1-\frac{2\epsilon}{e^{\eta}+ \lambda})\, ,
\end{equation}
where we have set $\overline{\kappa} = - \lambda = 1-\epsilon, \;
\overline{\kappa}^{-1}= -\lambda^{-1}=1+\epsilon $, which is
appropriate for the case of the F-type intermediate statistics.
Indeed many relations can be modified by the replacement
$\overline{\kappa} \rightarrow -\lambda$ to go from the B-type to
the F-type.

For the F-type intermediate statistics, the result as in
Eq.(\ref{46}) is still valid and consequently,  the mean
occupation number is given by the power series form exactly as in
Eq.(\ref{49}) except that now we have  $y=e^{\eta}+\lambda$ for
the F-type anyons. The approximate form for the occupation number
is thus
\begin{equation}\label{54}
    n\approx \frac{1}{e^{\eta}+ \lambda}\,.
\end{equation}
The detailed thermodynamic properties stemming from this form,
such as the equation of state, virial expansion etc.  for the
F-type intermediate statistics are as described in Ref.
\cite{RA-PNS1}.

It is also observed that the equality of the specific heats of
B-type and F-type intermediate statistics particles \cite{May}
also prevails as shown in Ref. \cite{RA-PNS1}, if we utilize the
approximate forms for the occupation numbers. This is a very
interesting result and may be true more generally for the exact
forms of the occupation numbers formulated in the present work.

\section{ The occupation number as an infinite continued fraction}

It is possible to obtain an expression for the mean occupation
number in terms of the infinite continued fraction (CF). Let us
begin with the series expression which may be expressed
conveniently in the form

\begin{equation}\label{55}
n=\frac{\alpha_1}{y} +\frac{\alpha_2}{y^2}+ \frac{\alpha_3}{y^3} +
\cdots \, ,
\end{equation}

\noindent where we have set $y=e^{\eta}-\overline{\kappa}$ and
$\alpha_1, \alpha_2$ etc. are determined from the previous
sections, specifically Eq.(\ref{49}), such as
\begin{eqnarray}\label{56}
    &&\alpha_1=1+\frac{1}{6}\,
\left(\epsilon^2+\epsilon^3+\frac{29}{30}\,\epsilon^4+\cdots\right)\,,
\nonumber\\
&&\alpha_2=\frac{\overline{\kappa}}{2}\,
    \left(\epsilon^2+\epsilon^3+\frac{7}{6}\,\epsilon^4+\cdots\right) \, ,
    \nonumber\\
&&\alpha_3=\frac{1}{3}\,\left(\epsilon^2+
\epsilon^3+\frac{13}{6}\,\epsilon^4+\cdots\right) \, ,
\end{eqnarray}
etc. by combining terms containing various powers of $\epsilon$ in
Eq.(\ref{49}).  There is a standard method by which this infinite
series can be put in the form of CF. The method of determining the
CF form of a function given by an infinite series is well-known in
the literature \cite{Wall,Andrews}.

We shall briefly summarize the procedure here. The general
continued fraction of order $r$ is of the form
\begin{equation}\label{60}
    C_r=  b_0+ {a_1\over \displaystyle b_1 + {\strut a_2 \over
    \displaystyle
     b_2 + {\strut a_3 \over \displaystyle b_3 + \cdots }}}\, ,
\end{equation}
where the constants $b_0, b_1, \cdots \; , a_0, a_1, \cdots $ can
be determined by a straightforward procedure. The various
convergents are $C_0, C_1, \cdots $ corresponding to
$r=0,1,2,\cdots, \infty$. Accordingly we have
\begin{eqnarray}\label{60a}
    &&C_0=b_0=: A_0/B_0 \, ; \nonumber \\
    &&C_1=b_0 + a_1/b_1= \frac{b_0b_1 + a_1}{b_1}=: \frac{A_1}{B_1}
    \,; \nonumber\\
&&C_2=b_0 + a_1/(b_1 + a_2/b_2)= b_0 + a_1b_2/(b_1 b_2 +
    a_2)=:\frac{A_2}{B_2}\, ,
\end{eqnarray}
etc. The parameters $A_n, B_n$ satisfy the two-term recurrence
relations \cite{Andrews}:
\begin{eqnarray}\label{60b}
    &&A_n=b_n A_{n-1}+ a_n A_{n-2}; \; A_{-1}=1 \, ;\nonumber\\
    &&B_n= b_n B_{n-1} + a_n B_{n-2}; \; B_{-1}=0\, .
\end{eqnarray}
By solving the recurrence relations,  the general CF can be
determined. We may quote two examples of this procedure. The
standard sine series may be expressed in the form of a CF as:
\begin{equation}\label{60c}
    \sin x=   {x\over \displaystyle 1 + {\strut x^2 \over
    \displaystyle
     2\cdot 3 - {\strut x^2 + 2\cdot 3 x^2\over \displaystyle 4 \cdot 5 -x^2 + \cdots }}}\,
     .
\end{equation}
Furthermore, we can also deal with the inverse problem, i.e.,
Given the standard series form of the cosine function,
\begin{equation}\label{60d}
    \cos x = 1- x^2/2! + x^4/4! + \cdots \, ,
\end{equation}
we can employ the above procedure and obtain the CF form for the
cosine function as:
\begin{equation}\label{60e}
    \cos x=   {1\over \displaystyle 1 + {\strut x^2 \over
    \displaystyle
     2\cdot 1 - {\strut x^2 + 2 x^2\over \displaystyle 4 \cdot 3 -x^2 + \cdots }}}\,
     .
\end{equation}
 Employing this procedure for our present problem, after some
algebra, the final result can be expressed by accordingly
obtaining  various convergents (approximants):
\begin{eqnarray}\label{57}
n_1&=& \frac{\alpha_1}{y}\, , \\
\label{58} n_2&=& -\frac{\alpha_2 \, y}{\alpha_1 \, y + \alpha_2}
\,
, \\
\label{59} n_3&=& -\frac{\alpha_3\, y}{\alpha_2 \, y + \alpha_3}
\, ,
\end{eqnarray}
etc.\\
In the literature on CF, the convergents, which may be obtained in a
straightforward manner after some algebra, play an important role.

The general form of the CF is given by the form $C_r$ as in
Eq.(\ref{60}) and the procedure  can be extended to many
convergents. The meaning of the convergent or the approximant is
evident.

Now the question which might arise is: what is the advantage of
CF? Other than the elegant mathematical form, we remark that there
is a distinct advantage. The Pade approximant is a well-known
application. Moreover there is a theorem \cite{Wall}, involving
the convergents $n_1, n_2, n_3 \cdots$ which may be stated as:
\begin{equation}\label{61a}
    n_1 < n_3 < \cdots <n   \;\;\;{\rm and}\;\;\;
    n_2 > n_4 . \cdots > n\, .
\end{equation}
 This immediately provides a
clarifying definition of successive approximations i.e., the above
inequality tells us how to obtain successive approximations of the
quantity $n$. Indeed, the above tells us immediately that the
exact form of $n$ lies between $n_1$ and $n_2$, hence its
importance. We can thus establish that the exact $n$ is bigger
than the first convergent $n_1= \alpha_1 / y $ but smaller than
$n_2$ obtained above.

\section{Summary and conclusion}

We have presented our formulation which leads to the conclusion
that the intermediate statistics, defined in terms of
$\kappa$-permutation, can be viewed as strictly connection with
the deformed algebra. We have been able to deduce this feature in
two steps: first, in Section 2, we have found that the exchange
symmetry characterizing the intermediate statistics can be
expressed in terms of basic numbers; secondly, in Section 3, we
have seen that the basic numbers naturally lead to the deformed
algebra. Although the quoted connection between
$\kappa$-permutation and deformed algebra cannot be supported at
this stage by a close mathematical demonstration, in our opinion,
the above results contribute to a deeper insight into the open
question related to the physical interpretation of the deformed
quantum algebra.

 Moreover, in order to describe intermediate statistics, continuously
interpolating anywhere between BE and FD statistics, we have
introduced the parameter $\kappa$ as a variable parameter, the
statistics determining parameter, so that $\kappa=+1,-1$ represent
the extreme values which correspond to the standard BE and FD
statistics (boson and fermion oscillator algebra) in the limits.
Our formulation of the mean occupation number and other
thermodynamic parameters in terms of the infinite continued
fraction is a new feature, not known in the literature. Its
importance stems from the possibility of approximations i.e., the
validity of approximations in the theory.

The different behavior from the undeformed quantum theory can be
dealt with in the statistical behavior of a complex systems,
intrinsically contained in $q$-deformation, whose underlying
dynamics is spanned in many-body interactions and/or long-time
memory effects. This aspect is  outlined in several papers. Among
the others, for example, in Ref.\cite{svira} it has been shown
that $q$-deformation plays a significant role in understanding
higher-order effects in many-body nuclear interactions.

\end{document}